\documentclass[a4paper,english,8pt,amssymb,nofootinbib,superscriptaddress]{revtex4-2}
\usepackage{lmodern}
\usepackage{lmodern}
\usepackage[T1]{fontenc}
\usepackage[utf8]{inputenc}
\usepackage{babel}
\usepackage{verbatim}
\usepackage{bm}
\usepackage{amsmath}
\usepackage{amsthm}
\usepackage{amssymb}
\usepackage{esint}
\usepackage[unicode=true,pdfusetitle,
 bookmarks=true,bookmarksnumbered=false,bookmarksopen=false,
 breaklinks=false,pdfborder={0 0 1},backref=false,colorlinks=false]
 {hyperref}

\usepackage[svgnames]{xcolor}

\makeatletter

\pdfpageheight\paperheight
\pdfpagewidth\paperwidth

\usepackage{babel}
\usepackage{amsthm}\usepackage{bm}\usepackage{verbatim}\usepackage[most]{tcolorbox}

\usepackage{bm}

\makeatother

\begin{document}
\title{Almost Universal Metrics}
\author{Metin Gürses}
\email{gurses@fen.bilkent.edu.tr}

\affiliation{Department of Mathematics, Bilkent University,
06800 Ankara, Türkiye}
\author{Tahsin Çağrı Şişman}
\email{tahsin.c.sisman@gmail.com}

\affiliation{Department of Astronautical Engineering, University of Turkish Aeronautical
Association, 06790 Ankara, Türkiye}
\author{Bayram Tekin}
\email{bayram.tekin@bilkent.edu.tr}

\affiliation{Department of Physics, Bilkent University, 06800 Ankara, Türkiye}
\begin{abstract}
A metric is called \emph{universal} if every symmetric conserved rank-two tensor constructed locally from the metric, the curvature, and covariant derivatives of the curvature is proportional to the metric.  Such a metric solves arbitrary metric-based higher-curvature field equations after, at most, a change in the effective cosmological constant. A metric is called \emph{almost universal} if the same field equations reduce instead to a finite set of algebraic conditions and a single linear scalar equation for a profile function. Universal and almost universal metrics are useful because they keep higher-curvature gravity under analytic control. Plane waves are a well-known example of universal metrics, while the Kerr-Schild-Kundt class of metrics including the $pp$-waves is almost universal. Here, we add a new member to this class of metrics and show that nonzero constant curvature $pp$-wave metrics are also almost universal. They reduce the generic gravity field equations to those of cosmological Einstein-Maxwell theory with null dust. The background of the $pp$-waves has the topology $\mathbb{R}^{1,1}\times S^{2}$ and provides the missing partner to the Nariai metric with ${\rm dS}^{2}\times S^{2}$ and the Bertotti-Robinson metric with ${\rm AdS}^{2}\times S^{2}$ topologies. These quantum-protected metrics are of clear interest. We exemplify our results by using the quadratic and cubic gravity theories.
\end{abstract}
\maketitle
\section{Introduction}
Higher-curvature contributions, built from the Riemann tensor and its covariant derivatives, are expected to play a central role in any viable quantum theory of gravity. In particular, in the low-energy limit of string theory, the gravitational Lagrangian contains infinitely many curvature invariants and their derivatives. The corresponding field equations are typically extremely complicated and, in general, do \emph{not} admit generic Einstein metrics as solutions.

Nevertheless, certain special metrics in General Relativity retain an extraordinary property: they continue to solve the field equations of \emph{all} higher--derivative gravity theories constructed from curvature and its covariant derivatives, independently of the detailed form of the Lagrangian. To be precise, these metrics, dubbed \textit{universal} metrics, reduce any symmetric divergence-free two tensor $E_{\mu \nu}$ to a multiple of the metric as $E_{\mu \nu}= a g_{\mu \nu}$ where $a$ is a constant.  Understanding these ``quantum-protected'' universal metrics is therefore of considerable conceptual and technical interest.

Beginning with Deser \citep{deser} and subsequent works \citep{gib1}–\citep{hor}, plane waves have been shown to be completely insensitive to quantum corrections: all higher-order curvature contributions vanish identically in their effective field equations. For a long time, plane waves were believed to be the only universal metric. 

Both plane waves and $pp$-waves belong to the Kerr–Schild family and enjoy additional structural simplifications. The metrics to be studied here have analogous properties; let us recap some properties of the $pp$-waves and Kerr-Schild-Kundt (KSK) metrics. The $pp$-waves take their simplest form in the Brinkmann coordinates,
\begin{equation}
ds^{2}=2\,{\rm d} u\,{\rm d} v+{\rm d} x^{2}+{\rm d} y^{2}+2V(u,x,y)\,{\rm d} u^{2},\label{ks1}
\end{equation}
for which the Einstein field equations reduce to  $\Box V=0$, where $\Box$ denotes the Beltrami–Laplace operator of the flat background, that is $V=0$ in \eqref{ks1}. The plane waves correspond to the choice $V= S_{ab}(u)x^{a}x^{b}$ with $\mathrm{trace}\,S=0$; hence, the Einstein field equations are satisfied identically. Therefore, the plane wave metrics are said to be universal. A defining property of $pp$-waves is that the null vector $\lambda_{\mu}=\partial_{\mu}u$ is covariantly constant: $\nabla_{\mu}\lambda_{\nu}=0$. Such metrics do not automatically satisfy the field equations of a generic gravity theory; hence, they are not universal. However, they are \textit{almost universal} \citep{gib2,col1,gur1} since the field equations of a generic gravity theory reduce to,
\begin{equation}
\sum_{n=1}^{N}a_{n}\,\Box^{n}V=0.\label{cond1}
\end{equation}
where $a_{n}$ depend on the coupling constants of the chosen higher–curvature theory.

Beyond the flat spacetime, de Sitter (dS) and anti-de Sitter (AdS) spacetimes are also universal. This observation led to the generalization of almost universal metrics to KSK metrics \citep{gur1,gur4,gst3,gst5,ort1,ort4} with constant--curvature backgrounds, which take the form
\begin{equation}
g_{\mu\nu}=\bar{g}_{\mu\nu}+2V\,\lambda_{\mu}\lambda_{\nu},\label{un2}
\end{equation}
where $\bar{g}_{\mu\nu}$ is the (A)dS metric, and the null vector $\lambda_{\mu}$ satisfies
\begin{equation}
\nabla_{\mu}\lambda_{\nu}=\xi_{\mu}\lambda_{\nu}+\xi_{\nu}\lambda_{\mu},
\end{equation}
with $\xi_{\mu}$, defined by this equation, is orthogonal to $\lambda_{\mu}$; and $\lambda^{\mu}\partial_{\mu}V=0$. For the generic class of theories defined by
\begin{equation}
I=\int d^{D}x\,\sqrt{-g}\,f\!\left(g,\mathrm{Riem},\nabla \mathrm{Riem},\ldots\right),\label{eq:Generic_gravity_theories}
\end{equation}
where $f$ is a smooth function. The field equations of (\ref{eq:Generic_gravity_theories}) take the form
\begin{equation}
G_{\alpha\beta}+\Lambda_{0}g_{\alpha\beta}+H_{\alpha\beta}=\kappa T_{\alpha\beta},\label{eq:EoM_generic}
\end{equation}
where $G_{\alpha\beta}\equiv R_{\alpha\beta}-\frac{1}{2}g_{\alpha\beta}R$, and $H_{\alpha\beta}$ represents all the higher derivative terms. The KSK ansatz \eqref{un2}reduces the source-free field equations to: (i) algebraic relations among the couplings, the cosmological constant, and the background curvature, and (ii) a single linear partial differential equation for $V$ of the form
\begin{equation}
\sum_{n=1}^{N}a_{n}\,\Box^{n}V=0,\label{cond2}
\end{equation}
hence, they are almost universal.

Now, we move to the discussion of $pp$-wave metrics with non-flat transverse two-space. Recently, a new family of four-dimensional solutions of quadratic gravity was constructed in \citep{heefer} that are of the form
\begin{equation}
ds^{2}=2\,{\rm d}u\,{\rm d}v+h_{ab}(x^{c})\,{\rm d}x^{a}{\rm d}x^{b}+2V(u,x^{a})\,{\rm d}u^{2},\label{eq:pp-wave}
\end{equation}
where $(u,v)$ are null coordinates and $h_{ab}$ ($a=2,3$) is the metric of a two-dimensional constant-curvature surface $S^{2}$ or $H^{2}$. In particular, $h_{ab}$ depends only on $x^{a}$ and not on $u$ or $v$. These metrics do not solve Einstein's equations with a cosmological constant and are therefore genuinely non-Einsteinian $pp$-wave solutions of quadratic gravity. 

The layout of the paper is as follows: Section~II treats the product background and its Einstein--Maxwell interpretation.  Section~III adds the Kerr--Schild wave profile and derives the corresponding null-dust equation.   Sections~IV and V give two checks of the general reduction, first in quadratic gravity and then in a cubic curvature theory arising from string effective actions.

\section{Universal $\mathbb{R}^{1,1}\times S^{2}$ and $\mathbb{R}^{1,1}\times H^{2}$ Topologies}

Let us first focus on the corresponding background metric,
\begin{equation}
ds^{2}=2\,{\rm d}u\,{\rm d}v+h_{ab}\left(x^{c}\right)\,{\rm d}x^{a}\,{\rm d}x^{b},\label{backpp}
\end{equation}
to show that it is almost universal and solves the cosmological Einstein-Maxwell equations. Its Riemann tensor has a nontrivial part only in the transverse two-space:
\begin{equation}
\bar{R}_{abcd} =\frac{\bar{R}}{2}\left(h_{ac}h_{bd}-h_{ad}h_{bc}\right),\label{eq:Riembar_in_max_sym_const_curv_form}
\end{equation}
and therefore, the nonzero components of the Ricci and Einstein tensors are
\begin{equation}
\bar{R}_{ab} =\frac{\bar{R}}{2}\,h_{ab},\quad\bar{G}_{ab}  =\frac{\bar{R}}{2}\,\left(h_{ab}-g_{ab}\right),\quad \bar{R}\equiv 4\Lambda,\label{eq:Background_two-tensors}
\end{equation}
where we defined $\Lambda $ in the last equation in terms of the constant scalar curvature. With a null $\lambda_{\mu}=\delta_{\mu}^{0}$ satisfying  $\nabla_{\mu}\lambda_{\nu}=0$, the metric tensor can be written in the form
\begin{equation}
\bar{g}_{\mu\nu}=n_{\mu}\,\lambda_{\nu}+n_{\nu}\,\lambda_{\mu}+h_{\mu\nu},\label{eq:gbar_in_n_and_lambda}
\end{equation}
where $n_{\mu}=\delta_{\mu}^{1}$ satisfies $\nabla_{\mu}\,n_{\nu}=0$. Here, $h_{\mu\nu}$ has the non-zero components $h_{ab}$; therefore, it satisfies $\lambda^{\mu} h_{\mu\nu}=0$ and $n^{\mu} h_{\mu\nu}=0$. Let us show that $\bar{g}_{\mu\nu}$ is a solution to the cosmological Einstein-Maxwell theory with the action
\begin{equation}
I = \int d^{4}x\,\sqrt{-g}\left[\frac{1}{\kappa}\left(R-2\Lambda_{0}\right)-\frac{1}{4}F^{\mu\nu}F_{\mu\nu}\right].\label{eq:Einstein-Maxwell_action}
\end{equation}
 Assuming the field potential and its field strength tensor are as follows:
\begin{equation}
A_{\mu}=\varepsilon(u)\,v\lambda_{\mu}\Longrightarrow F_{\mu\nu}=\varepsilon(u)\left(n_{\mu}\,\lambda_{\nu}-n_{\nu}\,\lambda_{\mu}\right),\label{eq:Vector_field}
\end{equation}
representing an electric field in the null direction as $F_{01}=E=-\varepsilon(u)$. The corresponding energy-momentum tensor $T_{\mu\nu}^{(F)} \equiv F_{\mu}\,^{\sigma}\,F_{\nu\sigma}-\frac{1}{4}\,\bar{g}_{\mu\nu}\,F^{\rho\sigma}F_{\rho\sigma}$ becomes
\begin{equation}
\frac{T_{\mu\nu}^{(F)}}{\varepsilon^{2}}= \frac{\bar{g}_{\mu\nu}}{2}-n_{\mu}\,\lambda_{\nu}-n_{\nu}\,\lambda_{\mu} = h_{\mu\nu}-\frac{1}{2}\,\bar{g}_{\mu\nu}.\label{en2}
\end{equation}
since $\lambda^{\mu}\,n_{\mu}=1$ and  $n_{\mu}\,\lambda_{\nu}+n_{\nu}\lambda_{\mu}=\bar{g}_{\mu\nu}-h_{\mu\nu}$. Using this result, together with the Ricci tensor and scalar curvature from (\ref{eq:Background_two-tensors}) in the cosmological Einstein-Maxwell field equations $
G_{\mu\nu}+\Lambda_{0}g_{\mu\nu}=\kappa T_{\mu\nu}^{(F)}$, yields
\begin{equation}
\left(2\Lambda-\kappa\varepsilon^{2}\right)h_{\mu\nu}+\left(-2\Lambda+\Lambda_{0}+\frac{\kappa\varepsilon^{2}}{2}\right)\,\bar{g}_{\mu\nu}=0.
\end{equation}
Therefore, the algebraic equations
\begin{equation}
2\Lambda-\kappa\varepsilon^{2}  =0,\quad
-2\Lambda+\Lambda_{0}+\frac{\kappa\varepsilon^{2}}{2} =0,\label{eq:epsilon-Lambda_0_eqn}
\end{equation}
must be satisfied for a solution to exist. Note that only positive scalar curvature spacetimes are possible, and $\varepsilon$ must be a constant. Solving these algebraic equations yields the cosmological constant in terms of the electric field as
\begin{equation}
\Lambda=\Lambda_{0}=\kappa \frac{\varepsilon^{2}}{2}=\kappa \frac{E^{2}}{2}.
\end{equation}
Therefore, the electromagnetic energy density determines the cosmological constant.

With the topology  $\mathbb{R}^{1,1}\times S^{2}$, $\bar{g}_{\mu\nu}$ is a critical point between the Nariai solution ${\rm dS}^{2}\times S^{2}$ and the Bertotti-Robinson solution ${\rm AdS}^{2}\times S^{2}$; further properties of this metric were discussed in \citep{Gurses-Flat_branch}.

As a result, the background metric \eqref{backpp} satisfies the cosmological Einstein-Maxwell field equations for the electromagnetic four-potential
\begin{equation}
A_{\mu}=\sqrt{\frac{2\Lambda}{\kappa}}\,v\lambda_{\mu},
\end{equation}
where $\lambda_{\mu}=\partial_{\mu}u$, and $\Lambda$ is a positive cosmological constant.

In \citep{gur9}, we showed that \eqref{backpp} solves the generic gravity field equations (\ref{eq:EoM_generic}); hence, it is \emph{universal}. This property means that for the metric (\ref{backpp}), any rank-two symmetric tensor built from the metric, its curvature, and covariant derivatives is a linear combination of $\bar{g}_{\mu\nu}$ and $h_{\mu\nu}$. Therefore, for any generic gravity theory, $H_{\mu\nu}$ in \eqref{eq:EoM_generic} representing all the higher derivative terms takes the form
\begin{equation}\label{gen0}
H_{\mu\nu}=e_{0}\,\bar{g}_{\mu\nu}+e_{1}\,h_{\mu\nu},
\end{equation}
where $e_{0}$ and $e_{1}$ are constants that depend on the theory parameters and the cosmological constant $\Lambda$. Then, the field equations (\ref{eq:EoM_generic}) with the electro-magnetic source represented by the four-potential (\ref{eq:Vector_field}) reduce to
\begin{equation}
\left(2\Lambda-\kappa\varepsilon^{2}+e_{1}\right)h_{\mu\nu}+\left(-2\Lambda+\Lambda_{0}+\frac{\kappa\varepsilon^{2}}{2}+e_{0}\right)\bar{g}_{\mu\nu}=0.
\end{equation}
To have a solution, the algebraic equations
\begin{equation}
2\Lambda-\kappa\varepsilon^{2}+e_{1}  = 0,\quad
-2\Lambda+\Lambda_{0}+\frac{\kappa\varepsilon^{2}}{2}+e_{0}  = 0,\label{eq:epsilon_Lambda_f(Riem)}
\end{equation}
must be satisfied.

This reduction in the field equations suggests that for the metric (\ref{backpp}), the field equations of any generic gravity theory reduce to the Einstein-Maxwell field equations with a cosmological constant since the Einstein tensor given in (\ref{eq:Background_two-tensors}) with $\bar{R}=4\Lambda$, that is $G_{\mu\nu}=2\Lambda h_{\mu\nu}-2\Lambda\bar{g}_{\mu\nu}$, becomes
\begin{equation}
G_{\mu\nu}=\left(\kappa\varepsilon^{2}-e_{1}\right)h_{\mu\nu}-\left(\Lambda_{0}+\frac{\kappa\varepsilon^{2}}{2}+e_{0}\right)\bar{g}_{\mu\nu},
\end{equation}
where we used (\ref{eq:epsilon_Lambda_f(Riem)}). Then, writing $h_{\mu\nu}$ in terms of the energy-momentum tensor of an electromagnetic field $A_{\mu}=\Upsilon \,v\lambda_{\mu}$ as
\begin{equation}
h_{\mu\nu}=\frac{1}{\Upsilon^{2}}T_{\mu\nu}^{(F)}+\frac{1}{2}\bar{g}_{\mu\nu},
\end{equation}
yields
\begin{equation}
G_{\mu\nu}=\frac{\kappa\varepsilon^{2}-e_{1}}{\Upsilon^{2}}T_{\mu\nu}^{(F)}-\left(\Lambda_{0}+e_{0}+\frac{e_{1}}{2}\right)\bar{g}_{\mu\nu}.\label{eq:Einstein_1}
\end{equation}
This equation reduces to the Einstein-Maxwell field equations with updated parameters,
\begin{equation}
\Upsilon^{2}  \equiv\varepsilon^{2}-\frac{e_{1}}{\kappa},\quad
\bar{\Lambda}_{0} \equiv\Lambda_{0}+e_{0}+\frac{e_{1}}{2}.\label{eq:Upsilon-Lambdabar_defn}
\end{equation}
Then, to summarize, let the spacetime metric be given as \eqref{backpp}, then the field equations of any generic gravity theory reduce to the Einstein-Maxwell field equation with a cosmological constant,
\begin{equation}
G_{\mu\nu}+\bar{\Lambda}_{{\rm 0}}g_{\mu\nu}=\kappa T_{\mu\nu}^{(F)}(\Upsilon),
\end{equation}
where $\Upsilon$ and $\bar{\lambda}_0$ are defined with \eqref{eq:Upsilon-Lambdabar_defn}.

Let us show that the background metric (\ref{backpp}) is of type D. The Weyl tensor $\bar{C}_{\mu\alpha\nu\beta}$ is found to be
\begin{equation}
\bar{C}_{\mu\alpha\nu\beta}  =  \frac{\bar{R}}{2}\left(h_{\mu\nu}h_{\alpha\beta}-h_{\mu\beta}h_{\alpha\nu}\right)\nonumber +\frac{\bar{R}}{6}\left(\bar{g}_{\mu\nu}\bar{g}_{\alpha\beta}-\bar{g}_{\mu\beta}\bar{g}_{\alpha\nu}\right)
  -\frac{\bar{R}}{4}\left(\bar{g}_{\mu\nu}h_{\alpha\beta}+\bar{g}_{\alpha\beta}h_{\mu\nu}-\bar{g}_{\mu\beta}h_{\alpha\nu}-\bar{g}_{\alpha\nu}h_{\mu\beta}\right). \quad \label{eq:Cbar}    
  \end{equation}
Hence, the background spacetime is not conformally flat since $\bar{C}_{\mu\alpha\nu\beta}\neq 0$ for $\bar{R}\neq 0$. To show that  $\bar{C}_{\mu\alpha\nu\beta}$ is of type D, it must satisfy \citep{Stephani_et_al}
\begin{equation}
\bar{C}_{\mu\alpha\nu[\beta}\lambda_{\sigma]}\lambda^{\alpha}\lambda^{\nu}=0,\quad\bar{C}_{\mu\alpha\nu[\beta}n_{\sigma]}n^{\alpha}n^{\nu}=0.
\end{equation}
Calculating $\bar{C}_{\mu\alpha\nu\beta}\lambda^{\alpha}\lambda^{\nu}$ and $\bar{C}_{\mu\alpha\nu\beta}n^{\alpha}n^{\nu}$, one has
\begin{equation}
\bar{C}_{\mu\alpha\nu\beta}\lambda^{\alpha}\lambda^{\nu}=\frac{\bar{R}}{6}\,\lambda_{\mu}\lambda_{\beta}, \quad \bar{C}_{\mu\alpha\nu\beta}n^{\alpha}n^{\nu}=\frac{\bar{R}}{6}\,n_{\mu}n_{\beta},
\end{equation}
therefore,  $\bar{C}_{\mu\alpha\nu[\beta}\lambda_{\sigma]}\lambda^{\alpha}\lambda^{\nu}=0$ and $\bar{C}_{\mu\alpha\nu[\beta}n_{\sigma]}n^{\alpha}n^{\nu}=0$. Therefore, the metric
\begin{equation}
ds^{2}=2\,{\rm d}u\,{\rm d}v + h_{ab}(x^{c})\,{\rm d}x^{a}{\rm d}x^{b},\label{backpp-1}
\end{equation}
defines a Type-D spacetime.

\section{Almost Universal Constant Curvature $pp$-waves}

Here, we extend the discussion of the fact that $\bar{g}_{\mu\nu}$ reduces the generic gravity field equations to the cosmological Einstein-Maxwell field equations to the Kerr-Schild metric
\begin{equation}
g_{\mu\nu}=n_{\mu}\,\lambda_{\nu}+n_{\nu}\,\lambda_{\mu}+h_{\mu\nu}+2V(u,x^a)\lambda_{\mu}\,\lambda_{\nu},\label{met1}
\end{equation}
for which the generic gravity field equations \textit{reduce} to those of cosmological Einstein-Maxwell with null dust. For this, let us first show that the Kerr-Schild metric (\ref{met1}) \textit{satisfies} the cosmological Einstein-Maxwell field equations with null dust with the energy density $\Phi$ as 
\begin{equation}
G_{\mu\nu}+\Lambda_{0}g_{\mu\nu}=\kappa T_{\mu\nu}^{(F)}+\kappa T_{\mu\nu}^{(\Phi)}.\label{eq:EM-null_dust_EoM}
\end{equation}
We show this result starting with the Einstein tensor of (\ref{met1}), which has the form
\begin{equation}
 G_{\mu\nu}=\frac{\bar{R}}{2}\,\left(h_{\mu\nu}-g_{\mu\nu}\right)\,-\,\lambda_{\mu}\,\lambda_{\nu}\,\Box V, \label{ein1}
\end{equation}
where we used  (\ref{eq:Background_two-tensors}). As in the background case, for the vector field $ A_{\mu}=\varepsilon\,v\lambda_{\mu}$, $F_{\mu}\,^{\sigma}$ takes the form $F_{\mu}\,^{\sigma}=\varepsilon\left(n_{\mu}\,\lambda^{\sigma}-\delta_{u}^{\sigma}\,\lambda_{\mu}+2V\lambda_{\mu}\lambda^{\sigma}\right)$; therefore,
\begin{equation}
T_{\mu\nu}^{(F)}=\varepsilon^{2}\,\left(-\frac{1}{2}\,g_{\mu\nu}+h_{\mu\nu}\right).\label{en1}
\end{equation}
In addition, let the null dust with energy density $\Phi$ have the energy-momentum tensor $T_{\mu\nu}^{(\Phi)} = \Phi \lambda_{\mu} \lambda_{\nu}$. Now, putting (\ref{ein1}), (\ref{en1}), $T_{\mu\nu}^{(\Phi)} = \Phi \lambda_{\mu} \lambda_{\nu}$, and $R=\bar{R}=4 \Lambda$ in the field equation (\ref{eq:EM-null_dust_EoM}) yields
\begin{equation}
0 =  \left(2\Lambda-\kappa\varepsilon^{2}\right)h_{\mu\nu}+\left(-2\Lambda+\Lambda_{0}+\frac{\kappa\varepsilon^{2}}{2}\right) g_{\mu\nu} \nonumber 
-\left(\Box V+\kappa \Phi\right)\lambda_{\mu}\,\lambda_{\nu},
\end{equation}
hence the set of equations,
\begin{equation}
2\Lambda-\kappa\varepsilon^{2}  =0,\quad
-2\Lambda+\Lambda_{0}+\frac{\kappa\varepsilon^{2}}{2}  =0,\quad
\Box V+\kappa \Phi  =0, \label{eq:h-g-Null_coefs}
\end{equation}
must be satisfied. The first equation requires $\Lambda =\kappa \varepsilon^{2}/2$ which means that only positive scalar curvature is possible. As a result, the Kerr-Schild metric \eqref{eq:pp-wave} satisfies the cosmological Einstein-Maxwell field equations with a null dust, once the electromagnetic four-potential is
\begin{equation}
A_{\mu}=\sqrt{\frac{2\Lambda}{\kappa}}\,v\lambda_{\mu},
\end{equation}
where $\Lambda$ is a positive cosmological constant, and the null dust has the energy density
\begin{equation}
\Phi=-\frac{1}{\kappa}\square V.
\end{equation}

Now we follow the same procedure as we did for the metric $\bar{g}_{\mu\nu}$. Since the $pp$-wave metric with constant curvature (\ref{eq:pp-wave}) is an almost universal metric, as shown in \citep{gur9}, the higher curvature contributions to the field equations become
\begin{equation}\label{gen1}
H_{\mu\nu}=e_{0}\,g_{\mu\nu}+e_{1}\,h_{\mu\nu}+\lambda_{\mu}\,\lambda_{\nu}\sum_{n=1}^{N}\,c_{n}\square^{n}V,
\end{equation}
where $e_{0}$, $e_1$ and $c_{n}$ are constants that depend on the theory parameters and the cosmological constant $\Lambda$. In addition, $N$ represents the highest derivative order in the field equations of the theory. Then, the generic gravity field equations with electromagnetic and null dust sources,
\begin{equation}
G_{\mu\nu}+\Lambda_{0}g_{\mu\nu}+H_{\mu\nu}=\kappa T_{\mu\nu}^{(F)}+\kappa T_{\mu\nu}^{(\Phi)},\label{eq:EM-null_dust_generic_EoM}
\end{equation}
 reduce to
\begin{equation}
0=  \left(2\Lambda-\kappa\varepsilon^{2}+e_{1}\right)h_{\mu\nu}+\left(-2\Lambda+\Lambda_{0}+\frac{\kappa\varepsilon^{2}}{2}+e_{0}\right)g_{\mu\nu}\nonumber
  -\left(\Box V+\kappa\Phi-\sum_{n=1}^{N}c_{n}\square^{n}V\right)\lambda_{\mu}\lambda_{\nu}, 
\end{equation}
once the electro-magnetic source is represented with the four-potential (\ref{eq:Vector_field}). Therefore, the set of equations
\begin{align}
0 & =2\Lambda-\kappa\varepsilon^{2}+e_{1},\quad  0=-2\Lambda+\Lambda_{0}+\frac{\kappa\varepsilon^{2}}{2}+e_{0}, \label{eq:h_coef-g_coef_f(Riem)}\\
0 & =\Box V+\kappa \Phi-\sum_{n=1}^{N}c_{n}\square^{n}V, \label{eq:Null_coef_f(Riem)}
\end{align}
must be satisfied to have a solution.

This reduction in the field equations suggests that for the metric (\ref{eq:pp-wave}), the field equations of any generic gravity theory are equivalent to the cosmological Einstein-Maxwell field equations with null dust. To see this, first, using $\bar{R}=4\Lambda$ in the Einstein tensor (\ref{ein1}), and then using  (\ref{eq:h_coef-g_coef_f(Riem)}) and (\ref{eq:Null_coef_f(Riem)}) yields
\begin{equation}
G_{\mu\nu}=  \left(\kappa\varepsilon^{2}-e_{1}\right)h_{\mu\nu}-\left(\Lambda_{0}+\frac{\kappa\varepsilon^{2}}{2}+e_{0}\right)g_{\mu\nu}\nonumber  +\left(\kappa\Phi-\sum_{n=1}^{N}c_{n}\square^{n}V\right)\lambda_{\mu}\lambda_{\nu}.
\end{equation}
Then, writing $h_{\mu\nu}$ in terms of the energy-momentum tensor of an electromagnetic field $A_{\mu}=\Upsilon\,v\lambda_{\mu}$ as
\begin{equation}
h_{\mu\nu}=\frac{1}{\Upsilon^{2}}T_{\mu\nu}^{(F)}+\frac{1}{2} g_{\mu\nu},
\end{equation}
yields
\begin{equation}
G_{\mu\nu}=  \frac{\kappa\varepsilon^{2}-e_{1}}{\Upsilon^{2}}T_{\mu\nu}^{(F)}-\left(\Lambda_{0}+e_{0}+\frac{e_{1}}{2}\right)g_{\mu\nu}\nonumber  +\left(\kappa\Phi-\sum_{n=1}^{N}c_{n}\square^{n}V\right)\lambda_{\mu}\lambda_{\nu}.\label{eq:Einstein_1_f(Riem)}
\end{equation}
Finally, with the definitions,
\begin{equation}
\Upsilon^{2}  \equiv\varepsilon^{2}-\frac{e_{1}}{\kappa},\quad
\bar{\Lambda}_{0} \equiv\Lambda_{0}+e_{0}+\frac{e_{1}}{2},\quad
\bar{\Phi}  \equiv\Phi-\frac{1}{\kappa}\sum_{n=1}^{N}c_{n}\square^{n}V,\label{eq:beta_defn_LambdaBar_defn_PhiBar_defn}
\end{equation}
(\ref{eq:Einstein_1_f(Riem)}) reduces to
\begin{equation}
G_{\mu\nu}+\bar{\Lambda}_{{\rm 0}}g_{\mu\nu}=\kappa T_{\mu\nu}^{(F)}+\kappa T_{\mu\nu}^{(\bar{\Phi})},
\end{equation}
which are the cosmological Einstein-Maxwell field equations with null dust in terms of the updated parameters. 

To summarize, let the spacetime metric be given as the $pp$-wave metric with constant curvature  \eqref{eq:pp-wave}, then the field equations of any generic gravity theory reduce to the cosmological Einstein-Maxwell field equation with a null dust fluid.

We now show that the constant-curvature $pp$-wave spacetimes \eqref{met1} are of Petrov type~II. First, note that the Weyl tensor $C_{\mu\alpha\nu\beta}$ can be written as
\begin{equation}
C_{\mu\alpha\nu\beta}=\bar{C}_{\mu\alpha\nu\beta}+4\lambda_{[\mu}\Omega_{\alpha][\beta}\lambda_{\nu]},\label{eq:Weyl_tensor}
\end{equation}
where $\Omega_{\alpha\beta}$ is defined as
\begin{equation}
\Omega_{\alpha\beta}\equiv-\left[\nabla_{\beta}\nabla_{\alpha}-\frac{1}{2}h_{\alpha\beta}\left(\square-\frac{\bar{R}}{3}\right)\right]V.\label{eq:Omega}
\end{equation}
Since $\lambda^{\mu}h_{\mu\nu}=0$, $\lambda^{\mu}\partial_{\mu}V=0$ and $\nabla_{\mu}\lambda_{\nu}=0$, we have $\lambda^{\mu}\Omega_{\mu\nu}=0$. Then, it follows that $C_{\mu\alpha\nu\beta}\lambda^{\beta}=\bar{C}_{\mu\alpha\nu\beta}\lambda^{\beta}$ which yields
\begin{equation}
C_{\mu\alpha\nu[\beta}\lambda_{\sigma]}\lambda^{\alpha}\lambda^{\nu}=\bar{C}_{\mu\alpha\nu[\beta}\lambda_{\sigma]}\lambda^{\alpha}\lambda^{\nu}=0.\label{eq:Type_II_cond}
\end{equation}
On the other hand, the $n_{\mu}=\delta_{\mu}^{v}$ one-form, which is null with respect to the background spacetime, now has the norm $n_{\mu}n^{\mu}=-2V$. Therefore, we need to define the other null vector with respect to the constant-curvature $pp$-wave spacetime as $N_{\mu}\equiv n_{\mu}+V\lambda_{\mu}$ for which $C_{\mu\alpha\nu[\beta}N_{\sigma]}N^{\alpha}N^{\nu}\ne0$ since $V$ has the coordinate dependence $V=V\left(u,x^{a}\right)$. Hence, the constant-curvature $pp$-wave spacetime is type II, as (\ref{eq:Type_II_cond}) is satisfied \cite{Stephani_et_al,Jordan_et_al-1961,Jordan_et_al-GRG}.

As a result, the Kerr-Schild metric \eqref{eq:pp-wave} is of type~II.

One can proceed without the null dust, in which case, the generic gravity field equations we have to solve 
\begin{equation}
0=\sum_{n=1}^{N}c_{n}\Box^{n}V=c_N \prod_{k=1}^{N-1}(\Box-m_{k}^{2})\,\Box V,\label{kok1}
\end{equation}
where we have factorized the equation in the second equality. Here,  $m_{k}^{2}$ are real or complex constants determined in terms of $c_n$. The general solution of (\ref{kok1}) can then be written as
\begin{equation}
V=\sum_{k=1}^{N-1}V_{k}+V_{N},\quad (\Box-m_{k}^{2})\,V_{k}  =0,\quad \Box V_{N}  =0,     
\end{equation}
where $m_{k}$'s are distinct and nonzero for $k=1,2,\cdots, N-1$. Note that if some $m_k$'s coincide or are zero, then new solutions arise as the second and third equations change. Thus, the single higher order equation \eqref{kok1} is equivalent to a set of $N$ Klein--Gordon--type equations for the components $V_{k}$  \cite{gur1,gur4,gst3,gur9,gur10}.

\section{Quadratic gravity}
As an illustration, consider the four-dimensional action of quadratic gravity with Maxwell electrodynamics,
\begin{equation}
\mathcal{L}=\frac{1}{\kappa}\left(R-2\Lambda_{0}+\alpha R^{2}+\beta R_{\mu\nu}R^{\mu\nu}\right) -\frac{1}{4}F^{\mu\nu}F_{\mu\nu},
\end{equation}
with constants $\alpha$, $\beta$, and the bare cosmological constant $\Lambda_0$. The quadratic curvature term $H_{\mu\nu}$ in the field equations was given in \cite{Stelle} or in \cite{DeserTekin-PRL,DeserTekin-PRD} as
\begin{align}
&H_{\mu\nu}=  2\alpha R\left(R_{\mu\nu}-\frac{1}{4}g_{\mu\nu}R\right)+\left(2\alpha+\beta\right)\left(g_{\mu\nu}\square-\nabla_{\mu}\nabla_{\nu}\right)R\nonumber \\
 & +\beta\square\left(R_{\mu\nu}-\frac{1}{2}g_{\mu\nu}R\right)+2\beta\left(R_{\mu\sigma\nu\rho}-\frac{1}{4}g_{\mu\nu}R_{\sigma\rho}\right)R^{\sigma\rho}.
\end{align}
Let us study the solution of this theory, given the metric and electromagnetic field ansätze as (\ref{backpp} or (\ref{eq:gbar_in_n_and_lambda}) and (\ref{eq:Vector_field}), respectively. Using (\ref{eq:gbar_in_n_and_lambda}) in $H_{\mu\nu}$ yields
\begin{equation}
H_{\mu\nu}=8\left(2\alpha+\beta\right)\Lambda^{2}h_{\mu\nu}-4\left(2\alpha+\beta\right)\Lambda^{2}g_{\mu\nu},\label{eq:H_quad_grav}
\end{equation}
therefore,  $e_{0}  =-4\left(2\alpha+\beta\right)\Lambda^{2}$ and $e_1=-2 e_0$. With the ansätze (\ref{eq:gbar_in_n_and_lambda}) and (\ref{eq:Vector_field}), the field equations of any generic gravity theory reduce to the Einstein-Maxwell field equation with a cosmological constant as
\begin{equation}
G_{\mu\nu}+\bar{\Lambda}_{0}g_{\mu\nu}=\kappa T_{\mu\nu}^{(F)}\left(\Upsilon\right),
\end{equation}
where the updated parameters $\Upsilon$ and $\bar{\Lambda}_{0}$ defined in (\ref{eq:Upsilon-Lambdabar_defn}) must satisfy the algebraic equations (\ref{eq:epsilon-Lambda_0_eqn}). Using $e_{0}$ and $e_{1}$, we found above in (\ref{eq:epsilon-Lambda_0_eqn}) yields $\Lambda =\Lambda_{0}$ and
\begin{equation}
8\left(2\alpha+\beta\right)\Lambda^{2}+2\Lambda =\kappa\varepsilon^{2}.
\end{equation}
Here, note that in contrast to Einstein-Maxwell theory, the quadratic gravity augmented with Maxwell electrodynamics can have both positive and negative $\Lambda$, which allows for both topology $\mathbb{R}^{1,1}\times S^{2}$ and topology $\mathbb{R}^{1,1}\times H^{2}$ for the background metric. In the absence of the electromagnetic field, that is, setting the parameter $\varepsilon$ to zero, the sourceless quadratic gravity admits a constant curvature solution for $1/\Lambda=-4\left(2\alpha+\beta\right)$ and $\Lambda=0$, which contradicts the metric ansatz. On the other hand, with the presence of the electromagnetic field, there are two solutions with a cosmological constant
\begin{equation}
\Lambda=\frac{-1\pm\sqrt{1+8\kappa\varepsilon^{2}\left(2\alpha+\beta\right)}}{8\left(2\alpha+\beta\right)}\label{eq:Lambda_quad_grav-Maxwell},\quad 2\alpha+\beta\neq 0.
\end{equation}
Therefore, the metric (\ref{backpp}) is a solution of quadratic gravity with Maxwell electrodynamics, where the electromagnetic four potential is (\ref{eq:Vector_field}), and the cosmological constant (which is equal to the bare cosmological constant) is determined by the theory parameters and the electromagnetic field, as given in (\ref{eq:Lambda_quad_grav-Maxwell}).

Now, let us study the solutions of quadratic gravity for the metric ansatz (\ref{met1}). With this metric, $H_{\mu\nu}$ becomes
\begin{equation}
H_{\mu\nu}= 8\left(2\alpha+\beta\right)\Lambda^{2}h_{\mu\nu}-4\left(2\alpha+\beta\right)\Lambda^{2}g_{\mu\nu}\nonumber  -\lambda_{\mu}\lambda_{\mu}\left(4\left(2\alpha+\beta\right)\Lambda\bar{\square}V-\beta\bar{\square}^{2}V\right),
\end{equation}
therefore, $e_{0}$ and $e_{1}$ are the same as the background values we gave after \eqref{eq:H_quad_grav} and
\begin{equation}
c_{1} =-4\left(2\alpha+\beta\right)\Lambda,\quad c_{2} =\beta.
\end{equation}
The solutions for the sourceless quadratic gravity were studied in \cite{heefer}. On the other hand, once the electromagnetic and null dust sources are introduced via the electromagnetic vector field (\ref{eq:Vector_field}) and the null dust energy momentum tensor $T_{\mu\nu}^{(\Phi)}=\Phi\lambda_{\mu}\lambda_{\nu}$, respectively, the field equations for quadratic gravity with these sources can be found from
\begin{equation}
G_{\mu\nu}+\bar{\Lambda}_{0}g_{\mu\nu}=\kappa T_{\mu\nu}^{(F)}\left(\Upsilon\right)+\kappa T_{\mu\nu}^{(\bar{\Phi})},
\end{equation}
since the field equations of any generic gravity theory reduce to the cosmological Einstein-Maxwell field equations involving a null dust source with the metric and the electromagnetic vector field ansätze (\ref{met1}) and (\ref{eq:Vector_field}). Here, the updated parameters $\Upsilon$ and $\bar{\Lambda}_{0}$ are given as (\ref{eq:h_coef-g_coef_f(Riem)}), which are the same as the above background case. On the other hand, the updated energy density of null dust is defined in (\ref{eq:Null_coef_f(Riem)}). These updated parameters must satisfy the set of equations (\ref{eq:h-g-Null_coefs}). The first two algebraic equations are the same as in the background case; therefore, they yield the same solutions for the cosmological constant in terms of either the bare cosmological constant or the theory parameters and the vector potential parameter. On the other hand, the third equation becomes
\begin{equation}
\kappa\Phi=\left(\beta\bar{\square}-\left[1+4\left(2\alpha+\beta\right)\Lambda\right]\right)\bar{\square}V.\label{eq:Phi_quad_grav}
\end{equation}
This equation determines the energy density of the null dust. However, it allows for the absence of null dust if
$\bar{\square}V=0$ or $V=V_{1}+V_{2}$ where $\bar{\square}V_{1}=0$ and
\begin{equation}
\left(\beta\bar{\square}-\left[1+4\left(2\alpha+\beta\right)\Lambda\right]\right)V_{2}=0.
\end{equation}
Or, if $1+4(2\alpha+\beta)\, \Lambda=0$ then $\bar{\square}^2\,V=0$ if $\beta\neq0$.

Therefore, the constant curvature $pp$-wave metric is a solution of quadratic gravity with Maxwell electrodynamics, for which the electromagnetic four potential is (\ref{eq:Vector_field}), while the null dust has the energy density given by (\ref{eq:Phi_quad_grav}) for a given metric function $V$. The cosmological constant determining the constant curvature is equal to the bare cosmological constant, and both are determined by the theory parameters and the electromagnetic field as given in (\ref{eq:Lambda_quad_grav-Maxwell}).

\section{Cubic gravity generated by string theory} 
As a second example, we study the four-dimensional cubic curvature gravity theory with the Lagrangian density,
\begin{equation}
\kappa \mathcal{L}=R-2\Lambda_{0}-\frac{\kappa}{4}F^{\mu\nu}F_{\mu\nu}
+\frac{\left(\alpha^{\prime}\right)^{2}}{24}\left(
R_{\alpha\beta}^{\mu\nu}R_{\mu\nu}^{\gamma\lambda}R_{\gamma\lambda}^{\alpha\beta}-2R_{\nu\beta}^{\mu\alpha}R_{\mu\lambda}^{\nu\gamma}R_{\alpha\gamma}^{\beta\lambda}\right),
\end{equation}
augmented with Maxwell electrodynamics. The $H_{\mu\nu}$ term in the field equations coming from the cubic curvature part is
\begin{align}
H_{\mu\nu}=  \frac{\left(\alpha^{\prime}\right)^{2}}{24} & \Biggl[6g_{(\mu|\rho}\nabla^{\lambda}\nabla_{\sigma}\left(-2R_{|\nu)\beta}^{\rho\alpha}R_{\alpha\lambda}^{\beta\sigma}+R_{|\nu)\lambda}^{\alpha\beta}R_{\alpha\beta}^{\rho\sigma}\right)\nonumber  +3R_{\rho\sigma(\mu|}^{\phantom{\rho\sigma(\mu|}\lambda}\left(-2R_{|\nu)\beta}^{\rho\alpha}R_{\alpha\lambda}^{\beta\sigma}+R_{|\nu)\lambda}^{\alpha\beta}R_{\alpha\beta}^{\rho\sigma}\right)\nonumber \\
 & -\frac{1}{2}g_{\mu\nu}\left(-2R_{\sigma\beta}^{\rho\alpha}R_{\rho\lambda}^{\sigma\gamma}R_{\alpha\gamma}^{\beta\lambda}+R_{\alpha\beta}^{\rho\sigma}R_{\rho\sigma}^{\gamma\lambda}R_{\gamma\lambda}^{\alpha\beta}\right)\Biggr].\label{eq:H_tensor_for_string_cubic_terms}
\end{align}This cubic gravity was obtained by using the three- and four-point scattering amplitudes of bosonic strings in \cite{Tseytlin}. In addition, we introduced the bare cosmological constant $\Lambda_{0}$. 

Let us study the solution of this theory given the metric and electromagnetic field ansätze as (\ref{met1}) and (\ref{eq:Vector_field}), respectively. Using (\ref{met1}) in $H_{\mu\nu}$ yields
\begin{equation}
H_{\mu\nu}= 6\left(\alpha^{\prime}\right)^{2}\Lambda^{3}h_{\mu\nu}-2\left(\alpha^{\prime}\right)^{2}\Lambda^{3}g_{\mu\nu}\nonumber-2\left(\alpha^{\prime}\right)^{2}\Lambda^{2}\lambda_{\mu}\lambda_{\nu}\bar{\square}V,
\end{equation}
therefore, we have
\begin{align}
e_{0}  =-2\left(\alpha^{\prime}\right)^{2}\Lambda^{3},\,\, 
e_{1} = -3 e_0,\,\,
c_{1} =-2\left(\alpha^{\prime}\right)^{2}\Lambda^{2}.
\end{align}
With the ansätze (\ref{met1}) and (\ref{eq:Vector_field}), the field equations of any generic gravity theory reduce to the cosmological Einstein-Maxwell field equation involving a null dust source, with the updated parameters $\Upsilon$, $\bar{\Lambda}_{0}$ and $\bar{\Phi}$ are given as (\ref{eq:h_coef-g_coef_f(Riem)}), and (\ref{eq:Null_coef_f(Riem)}). These updated parameters must satisfy the set of equations (\ref{eq:h-g-Null_coefs}). The first two algebraic equations become:
\begin{equation}
\varepsilon^{2} =\frac{2\Lambda}{\kappa}\left(1+2\left(\alpha^{\prime}\right)^{2}\Lambda^{2}\right),\quad
\Lambda_{0} =\Lambda\left(1-\left(\alpha^{\prime}\right)^{2}\Lambda^{2}\right),
\end{equation}
determining $\varepsilon$ and $\Lambda_{0}$ in terms of the cosmological constant. Here, note that $\Lambda$ needs to be positive since the left-hand side and the term in parentheses are positive in the first equation. In addition, in the absence of the electromagnetic source, there is no solution for the first equation except $\Lambda=0$ again, since the term in parentheses is positive, and this $\Lambda=0$ solution contradicts the constant curvature $pp$-wave ansatz. On the other hand, the third equation becomes
\begin{equation}
\Phi=-\frac{1}{\kappa}\left(1+2\left(\alpha^{\prime}\right)^{2}\Lambda^{2}\right)\bar{\Box}V,
\end{equation}
determining the energy density of the null dust. However, it allows the absence of null dust if $\bar{\square}V=0$.

\section{Conclusions} 
We have constructed a non-Einsteinian class of almost universal $pp$-wave metrics with constant-curvature transverse two-spaces.  The central point is not merely that these metrics provide new exact solutions, but that their curvature algebra closes on a small tensor basis.  For the product background $\mathbb{R}^{1,1}\times \Sigma_{2}$, with $\Sigma_{2}=S^{2}$ or $H^{2}$, every higher-curvature contribution to the metric field equations reduces to a linear combination of $g_{\mu\nu}$ and $h_{\mu\nu}$.  For the Kerr--Schild deformation, the only additional structure is $\lambda_{\mu}\lambda_{\nu}$ multiplied by a linear scalar operator acting on the profile $V$.

Consequently, generic metric-based higher-curvature theories reduce on this class to cosmological Einstein--Maxwell theory with renormalized parameters, and, for the wave deformation, with an additional null-dust source.  In the absence of null dust, the profile equation factorizes into a finite sequence of Klein--Gordon--type equations.  We verified the reduction explicitly in quadratic gravity and in the cubic curvature theory arising from string effective actions.

The result shows that almost universality does not require the background to be Einstein or maximally symmetric.  The product background considered here supplies a non-Einsteinian partner to the familiar plane-wave, $pp$-wave, and Kerr--Schild--Kundt examples, and it gives a tractable exact setting for testing higher-curvature corrections in gravitational effective theories.

Here, we have worked in four dimensions, it is straightforward to extend these results to $(D+1)$ dimensions, where the two-sphere $S^{2}$ or the two-dimensional hyperbolic space $H^2$ are replaced by a $(D-1)$--dimensional constant--curvature space $S^{D-1}$ or $H^{D-1}$, respectively. The proof follows closely the arguments developed for KSK metrics \citep{gur1}--\citep{gur4,gst3,gst5}; see also the forthcoming work \citep{gur9}.

\end{document}